\documentclass[12pt]{article}
\usepackage{graphicx}
\usepackage{epsfig}
\usepackage{color}
\begin{document}

\title{Discrete Holomorphicity at Two-Dimensional Critical Points\footnote
{Invited talk at the 100th Statistical Mechanics Meeting, Rutgers,
December 2008}}

\author{John Cardy\\
Rudolf Peierls Centre for Theoretical Physics\\
         1 Keble Road, Oxford OX1 3NP,
         U.K.\footnote{Address for correspondence.}\\
and All Souls College, Oxford.}

\date{July 2009}

\maketitle

\begin{abstract}
After a brief review of the historical role of analyticity in the
study of critical phenomena, an account is given of recent
discoveries of discretely holomorphic observables in critical
two-dimensional lattice models. These are objects whose
correlation functions satisfy a discrete version of the
Cauchy-Riemann relations. Their existence appears to have a deep
relation with the integrability of the model, and they are
presumably the lattice versions of the truly holomorphic
observables appearing in the conformal field theory (CFT)
describing the continuum limit. This hypothesis sheds light on the
connection between CFT and integrability, and, if verified, can
also be used to prove that the scaling limit of certain discrete
curves in these models is described by Schramm-Loewner evolution
(SLE).
\end{abstract}

\section{Introduction}
\label{intro} Analyticity has played a key role in the development
of our mathematical understanding of the nature of critical
phenomena. At various levels, the assertion of the analytic
dependence of a suitable physical observable on some variable has
provided a powerful starting point for theories, and its ultimate
failure has informed the next step in their refinement.

The first example of this was perhaps Landau theory, in which the
free energy $F[{\bf M}]$ is postulated to be analytic in the order
parameter $\bf M$:
$$
F[{\bf M}]=\int\left((\nabla{\bf M})^2+{\rm Tr}({\bf H}\cdot{\bf
M})+r_0{\rm Tr}{\bf M}^2+\lambda_3{\rm Tr}{\bf
M}^3+\cdots\right)d^d\!x\,.
$$
In this expression, all terms allowed by symmetry are in principal
present. The requirement that $\bf M$ should minimize $F$, $\delta
F/\delta{\bf M}=0$, implies that there are critical points (in the
mathematical sense) at certain values of the parameters. The
singular behavior close to these is described by the mathematical
theory of bifurcations. The critical exponents which emerge from
this analysis are super-universal in that they do not depend on
the spatial dimension $d$ or the symmetry of the order parameter.

Of course, the successes and the limitations of Landau theory are
well understood. It does not account for the fluctuations in the
order parameter. In high enough dimensions these can usually be
ignored as far as the universal behavior is concerned, but in
general they should be taken into account by computing the full
Landau-Ginzburg-Wilson partition function
$$
Z[{\bf H}]=\int_{|{\bf q}|<\Lambda}[d{\bf M}(x)]\,e^{-F[{\bf
M}]}\,,
$$
with a suitable cut-off $\Lambda$ on the short wavelength modes.
This of course is impossible except in some simple cases, but
these can often be used as a starting point for a perturbative
expansion. However this approach generally fails near the critical
point.

The renormalization group (RG) of Wilson, Fisher and others was a
way of dealing with this problem, by considering the response to
changes in the cut-off $\Lambda\to \Lambda e^{-\ell}$ in such a
way as to move the parameters into a region where perturbation
theory is applicable. Requiring that the long-distance physics
remains the same then leads to the RG flow equations
$$
\frac{d\lambda_j}{d\ell}=-\Lambda\frac{\partial\lambda_j}{\partial\Lambda}=-\beta_j(\{\lambda\})
$$
for the parameters $\{\lambda_j\}$ in $F[{\bf M}]$. The critical
behavior is then controlled by the fixed points where
$\beta_j(\{\lambda\})=0$. One of the crucial assumptions of the RG
theory is that these functions themselves are analytic in the
$\{\lambda_j\}$ near the fixed point. Thus analyticity is moved to
a higher level of abstraction: the non-analyticity of the free
energy results from a potentially infinite number of applications
of the analytic RG mapping. [Interestingly enough there is very
little direct evidence for the strict validity of this assumption.
It is largely based on analysis of perturbation theory which begs
the question. Solvable models in two dimensions have so much
analytic structure built into them that analyticity of the RG
flows is almost inevitable, and possible detailed violations of
analyticity have not, to my knowledge, been carefully studied in
numerical simulations in higher dimensions. Indeed in one case,
$d=0$, when the beta-function can be evaluated analytically, it
displays an essential singularity at the fixed point.]

The next example of the use of analyticity is in the solution of
integrable models in two dimensions. In this case the independent
variable parameterizes the solution manifold on which the
Boltzmann weights satisfy the Yang-Baxter equations. Its physical
interpretation, to be discussed later, is the degree of anisotropy
of the model in two-dimensional euclidean space. The Yang-Baxter
equations imply that the row-to-row transfer matrices at different
values of this parameter commute. Assuming that the analytic
properties of the local weights in the parameter lift to
thermodynamic quantities, Baxter and others have shown that these
obey functional relations which often determine them completely.

The other main approach to understanding two-dimensional critical
behaviour attempts to describe the continuum critical scaling
limit directly, and also exploits analyticity. This is the
approach of conformal field theory (CFT), more recently also
linked to Schramm-Loewner evolution (SLE) \cite{sle}. In this case
the analyticity is in the two-dimensional coordinates $z=x+iy$ and
$\bar z=x-iy$. In CFT, correlation functions of certain
observables are holomorphic functions of $z$ (or antiholomorphic
functions of $\bar z$.) These observables are of two types:
conserved currents corresponding to continuous symmetries, such as
the stress tensor $\big(T(z),\overline T(\bar z)\big)$, or
so-called parafermions $\psi_s(z)$ whose 2-point correlation
function in the full plane have the form
$$
\langle\psi_s(z_1)\psi_s(z_2)\rangle\sim(z_2-z_2)^{-2s}\,,
$$
where $s$, the conformal spin, is in general fractional.

These holomorphic (and antiholomorphic) objects are the building
blocks of many CFTs. They also give an explicit meaning to the
statement of conformal covariance: if we consider the scaling
limit of a critical lattice model in a domain $\cal D$, then the
behavior of a multi-point holomorphic correlator is completely
determined by its singular behavior at coincident points and the
boundary conditions on $\partial{\cal D}$. If these are themselves
conformally covariant (\em e.g. \em if $\lim_{z\to\partial{\cal
D}}\arg\psi_s(z)$ is determined by the boundary tangent angle),
then, under any conformal mapping $f:{\cal D}\to{\cal D}'$ of the
interior of $\cal D$ to another domain ${\cal D}'$
$$
\langle\psi_s(z_1)\psi_s(z_2)\rangle_{\cal D}= f'(z_1)^s f'(z_2)^s
\langle\psi_s(z_1')\psi_s(z_2')\rangle_{{\cal D}'}\,.
$$

It is the purpose of this paper to review recent work which forges
a link between these last two realizations of analyticity. More
specifically, starting from certain lattice models we identify
so-called discretely holomorphic observables, whose correlators
satisfy a lattice version of the Cauchy-Riemann equations. These
have fractional spin by construction, and are presumably the
lattice precursors of the parafermions in the corresponding CFT.
We find, as expected, that discrete holomorphicity holds only when
the Boltzmann weights lie on the critical manifold of the model,
but, more surprisingly, that they also lie on the \em integrable
\em critical manifold, that is, they satisfy the Yang-Baxter
equations.

This work has been described in detail in several papers. In
\cite{cardyriva,smirnov} a discretely holomorphic observable was
identified for the random cluster representation of the $q$-state
Potts model. In \cite{cardyraj} several were found, directly from
the Boltzmann weights, for the $Z_N$ clock models, and the close
connection to integrability was first observed. Smirnov
\cite{smirnov} also found such an observable for the random curve
representation of the O$(n)$ model on the honeycomb lattice. This
was extended to the O$(n)$ model, and further generalizations, on
the square lattice in \cite{ikhlefcardy}, where the parameter
space is large enough to make clear the connection to
integrability. Smirnov's objective in identifying these
holomorphic objects associated with curves was to show that their
scaling limit in described by Schramm-Loewner evolution (SLE).
This approach is fully described in \cite{smirnov} and we refer
the reader there. Suffice it to say that proving that the
discretely holomorphic lattice observables go over into fully
holomorphic quantities in the scaling limit is in general a very
difficult problem which has been solved completely only in a few
special cases, in particular that of the Ising model ($q=2$ or
$n=1$) \cite{smirnov}.

In this account we shall describe in detail only two examples. The
first is the simplest: the case $N=2$ of the $Z_N$ model, more
usually known as the Ising model. In this case the parafermion has
conformal spin $s=\frac12$ and is related to the fermionic objects
used in the various exact solutions of the model, going back to
Onsager. It is defined, in a standard way, as a product of nearby
order and disorder variables. This identification dates back (at
least) to Fradkin and Kadanoff \cite{fradkin}, but the simple
argument given in \cite{cardyraj} that a particular lattice
definition is discretely holomorphic is, to our knowledge, new. An
essential ingredient in the definition of a disorder operator is
that the model must possess a dual description in the sense of
Kramers and Wannier. However, the second example is the O$(n)$
model which does not possess this. In this case the parafermionic
observables are instead defined in terms of the random curve
representation of the model, which is nonlocal in terms of the
original spin degrees of freedom.

\begin{figure}
\begin{center}
\includegraphics[width=3cm]{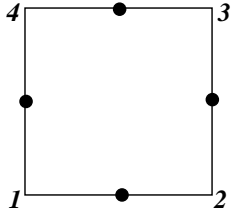}
\caption{A face of the square lattice. The discretely holomorphic
function $F(z)$ is defined at the mid-points of the edges.}
\label{fig1}
\end{center}
\end{figure}

Before proceeding, however, we give a proper definition of
discrete holomorphicity. Suppose $\cal G$ is a planar graph
embedded in ${\bf R}^2$, for example a square lattice. Let
$F(z_{ij})$ be a complex-valued function defined at the midpoints
$z_{ij}$ of each edge $(ij)$. Then $F$ is discretely holomorphic
on $\cal G$ if
\begin{equation}\label{holo}
\sum_{(ij)\in{\cal F}}F(z_{ij})(z_j-z_i)=0\,,
\end{equation}
where the sum is over the edges of each face $\cal F$ of $\cal G$.
This is a discrete version of the contour integral. For a square
lattice (see Fig.~\ref{fig1}) it reduces to
$$
F(z_{12})+iF(z_{23})+i^2F(z_{34})+i^3F(z_{41})=0\,.
$$
A little thought shows, however, that the total number of such
equations, one for each face of $\cal G$, is in general far less
than the number of unknowns, one for each edge. Therefore even on
the lattice this system does not determine $F(z_{ij})$ for
suitable boundary conditions, unless further information is
available. (For the Ising model it turns out that more is known
about the phase of $F$, so the system is rigid.)  However, in the
continuum limit as the lattice spacing tends to zero, we can
approximate arbitrarily closely the contour integral $\int_C
F(z)dz$ by a sum of the left-hand side of (\ref{holo}) over faces
$\cal F$ which tile the interior of $C$, and therefore assert that
is vanishes for all reasonable contours $C$. Morera's theorem then
assures us that \em if \em $F(z)$ is continuous then it is
analytic. In field theory the continuity of correlation functions
is usually taken for granted, but strictly this needs to be
proved.

\section{The Ising and $Z_N$ models}
\label{sec1} 
\begin{figure}
\begin{center}
\includegraphics[width=8cm]{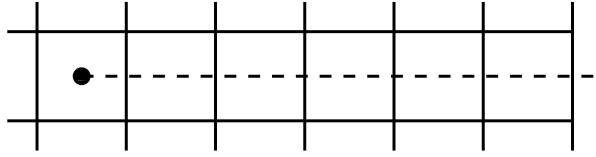}
\caption{The string defining a disorder operator.} \label{fig2}
\end{center}
\end{figure}

The square lattice Ising model has spins $s(r)=\pm1$
at the vertices $r$. The Boltzmann weights are $\exp(-{\cal H})$
with ${\cal H}=-\sum_{rr'}J_{rr'}s(r)s(r')$, where the sum is over
edges $(rr')$. Note that the weights can also be written\\
$\prod_{rr'}\big(1+(\tanh J_{rr'})s(r)s(r')\big)$. The insertion
of a disorder operator $\mu(R)$ at the dual vertex $R$ corresponds
to changing $J_{rr'}\to-J_{rr'}$ on all edges which cross a
`string' attaching $R$ to the boundary, see Fig.~\ref{fig2}.
That is
\begin{equation}
\label{eq1} \mu(R)=\prod_{(rr')\updownarrow{\rm
string}}\frac{1-(\tanh J_{rr'})s(r)s(r')} {1+(\tanh
J_{rr'})s(r)s(r')}\,.
\end{equation}
The correlator $\langle\mu(R_1)\mu(R_2)\rangle$ corresponds to
such a string connecting $R_1$ and $R_2$, and is invariant under
deformations of the string.

The parafermionic variables $\psi_s(rR)$ are defined on the
midpoints of each edge $(rR)$ connecting the vertex $r$ to a
neighboring dual vertex $R$, which form the covering lattice:
$$
\psi_s(rR)=s(r)\cdot\mu(R)\,e^{-is\theta(rR)}\,.
$$
Here $\theta(rR)$ is the angle that $(rR)$ makes with (say) the
positive $x$-axis, but we have to be careful about its
multivaluedness (see below).

Consider now an elementary square (see Fig.~\ref{fig3}).
\begin{figure}
\begin{center}
\includegraphics[width=6cm]{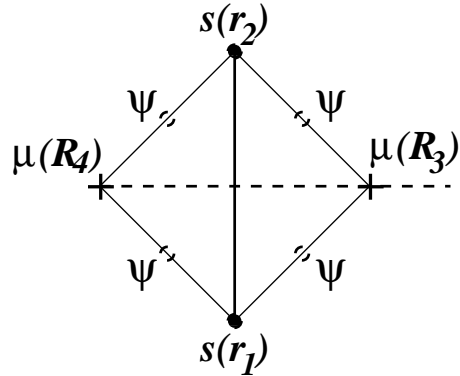}
\caption{Elementary square on whose edges the parafermionic
observables are defined.} \label{fig3}
\end{center}
\end{figure}
From (\ref{eq1}) we have
\begin{equation}\label{eq2}
\big(1+(\tanh J_y)s(r_1)s(r_2)\big)\,\mu(R_4)=\big(1-(\tanh
J_y)s(r_1)s(r_2)\big)\,\mu(R_3)\,,
\end{equation}
where we have set $J_{rr'}=J_{x,y}$ depending on whether $(rr')$
is parallel to the $x$ or $y$-axis. Now multiply (\ref{eq2}) by
$s(r_1)$ and $s(r_2)$ and use $s(r)^2=1$. This leads to two linear
equations in the four parafermionic variables defined on the edges
of the square. Simple algebra then shows that these imply the
discrete holomorphicity condition (\ref{holo}) as long as:
\begin{itemize}
\item we distort each square into a rhombus whose angle depends on
the anisotropy $J_y/J_x$;
\item we are careful, since $s$ is in general non-integer, to define $\theta(rR)$ consistently
so that it varies only by increments in the interval $(-\pi,\pi)$
on going around the square;
\item the couplings lie on the critical manifold $\sinh J_x\,\sinh
J_y=1$.
\end{itemize}

This example illustrates the general results discussed in the
introduction. It also shows how the lattice should be embedded in
${\bf R}^2$ so that its continuum limit is rotationally and
conformally invariant. However since the nearest neighbor Ising
model is integrable for all values of the couplings $(J_x,J_y)$,
it is not general enough to illustrate the role of integrability.

This is afforded by the generalization to the $Z_N$ models
\cite{cardyraj}. In these models the Ising spins are generalized
to complex roots of unity such that $s(r)^N=1$. The Boltzmann
weights take the form
$$
\prod_{rr'}\left(1+\sum_{k=1}^{N-1}x^{(k)}_{rr'}\big(s(r)^*s(r')\big)^k+{\rm
c.c.}\right)\,.
$$
It is then found that one can identify discretely holomorphic
parafermions, again the product of neighboring order and disorder
operators, with spins
$$
s=\frac{k(N-k)}N\qquad(1\leq k\leq N/2)\,,
$$
as long as the parameters lie on the manifold
\begin{equation}\label{eq:man}
x^{(k)}_x(\alpha)=\prod_{j=0}^{k-1}\frac{\sin\big((\pi
j+\alpha)/N\big)} {\sin\big((\pi(j+1)-\alpha)/N\big)}
\end{equation}
and $x^{(k)}_y(\alpha)=x^{(k)}_y(\pi/2-\alpha)$. Here $\alpha$ is
the half-angle at the vertex of the rhombus into which the square
lattice must be distorted, and therefore measures the degree of
anisotropy. However this manifold is precisely the critical \em
integrable \em case for the general nearest neighbor $Z_N$ model
found by Fateev and Zamolodchikov \cite{FZ1}. The values of the
conformal spins above agree precisely with those of the
parafermionic holomorphic conformal fields in the CFT postulated
by the same authors \cite{FZ2} to describe the continuum limit of
this model.

\begin{figure}
\begin{center}
\includegraphics[width=6cm]{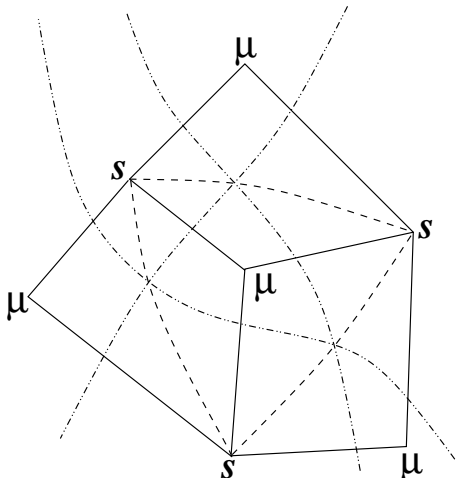}
\caption{Part of a Baxter lattice. The faces of the graph $\cal L$
formed by the curved lines are 2-colorable (not shown). Order
variables $s$ and disorder variables $\mu$ are associated with
alternately colored faces respectively. The covering lattice is
shown as solid lines. The theorem of Kenyon and Schlenker
\cite{kenyon} asserts that for every such graph the covering
lattice admits a rhombic embedding in the plane, that is one where
all its edges have the same length.} \label{baxter}
\end{center}
\end{figure}

\begin{figure}
\begin{center}
\includegraphics[width=8cm]{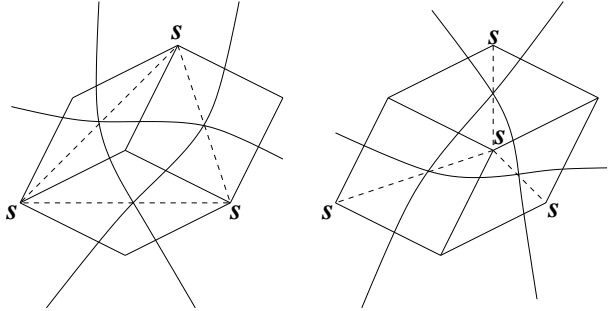}
\caption{Two different tilings of a hexagon by the same set of
three rhombi. The right-hand case has an additional vertex
associated to an order operator $s$ as compared to that on the
right. Discrete holomorphicity for each rhombus fixes the
couplings on the dashed lines to be related by the star-triangle
transformation. The two pictures are also related in the original
graph $\cal L$ by moving one of the curves past the vertex formed
by the other two -- the Yang-Baxter relation.}
\label{startriangle}
\end{center}
\end{figure}

The above analysis can be generalized simply to a general graph
$\cal L$ whose faces are 2-colorable, sometimes called a Baxter
lattice -- see Fig.~\ref{baxter}.
The $Z_N$ spins $s(r)$ are defined on the faces of a given color,
and the disorder variables $\mu(R)$ on the others. Neighboring
order and disorder operators then lie at the vertices of
quadrilaterals which tile the plane. A theorem due to Kenyon and
Schlenker \cite{kenyon} asserts that, under rather general
conditions, they can can be distorted in the plane so they are all
rhombi, that is, they form an isoradial lattice, on which nearest
neighbors are all the same distance apart. If now the interactions
across each rhombus are chosen to satisfy (\ref{eq:man}), where
$\alpha$ is the half-angle of the rhombus, then the associated
parafermion is discretely holomorphic on the isoradial lattice.
Moreover the corresponding weights satisfy the star-triangle, or
Yang-Baxter equations, as explained in Fig.~\ref{startriangle}.

We note that although we have defined the parafermions in this
case directly in terms of the modified Boltzmann weights, they may
also be identified with observables of random curves in the model.
For the case of the Ising model it is well known that the
high-temperature expansion, in powers of $\tanh J_{x,y}$, of the
correlation function $\langle s(r)s(0)\rangle$ can be expressed as
sum over graphs on the lattice. The argument is clearer for the
honeycomb lattice, when these graphs consist of non-intersecting
closed loops and one open curve $\gamma$ from $0$ to $r$. The
additional complications on the square lattice are believed to be
irrelevant to the scaling limit. The introduction of the disorder
operator $\mu(R)$ neighboring $s(r)$ acts to weight configurations
of the open curve with $(-1)^N$ where $N$ is the number of times
$\gamma$ crosses the string. This can also be written as
$e^{-is\theta_{0r}}$ where $\theta_{0r}$ is the winding angle of
$\gamma$. In the scaling limit this takes arbitrarily large values
and so it does not matter whether we compute it in increments of
$\pm 2\pi$ as it crosses the string, or simply in increments of
$\pm\frac\pi 2$ as we proceed along $\gamma$. This latter
definition corresponds to the one used by Smirnov \cite{smirnov}.

\section{The O$(n)$ model}\label{sec2}
\begin{figure}
  \begin{center}
    \scalebox{1}{\input{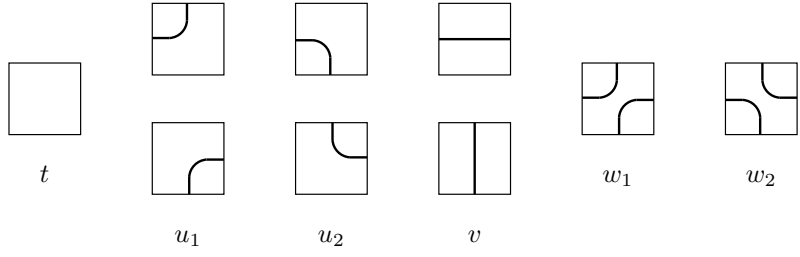}}
    \caption{Vertices of the O$(n)$ loop model on the square
    lattice.}
\label{fig4}
  \end{center}
\end{figure}

As a second example we consider the so-called O$(n)$ model on the
square lattice, first considered by Nienhuis \cite{nienhuis}.
Although it can be written in terms of $n$-component spins
$s_a(r)$ ($a=1,2,\ldots,n$) located on the edges of a square
lattice through local, but non-nearest neighbor, interactions, it
is more easily formulated in terms of a gas of dilute
non-intersecting planar loops. Each elementary square can take one
of the configurations shown in Fig.~\ref{fig4}, with the indicated
weights. When patched together these curves form closed loops,
each of which receives an additional weight $n$, which no longer
has to be an integer and may be parameterized by $n =-2\cos2\eta$
with $0\leq\eta\leq \frac\pi2$. Note that we have grouped the
anisotropic weights so that there is symmetry under reflections in
the diagonal axes. This symmetry is preserved when the plaquettes
are deformed into rhombi. Since every loop configuration has an
even number of plaquettes of type $u_1$ or $u_2$, the change
$(u_1,u_2) \to (-u_1,-u_2)$ does not affect the Boltzmann weights.

In order to define a suitable candidate for the parafermionic
observable we must consider not only closed loops but also open
curves which begin and end at different points, say $0$ and $r$.
Such configurations arise, as in the previous section, if we
compute the correlator $\langle s_a(r)s_a(0)\rangle$ of the O$(n)$
spins. Once again, to define an object with non-zero spin, we
additionally weight each open curve by a phase factor
$e^{-is\theta_{0r}}$, where $\theta_{0r}$ is the winding angle
from $0$ to $r$, the accumulation of the turns through
$\pm\frac\pi 2$. Note that this can be arbitrarily large. This
type of observable for the O$(n)$ model was first considered on
the honeycomb lattice by Smirnov \cite{smirnov}.

\begin{figure}
  \begin{center}
    \includegraphics[scale=1]{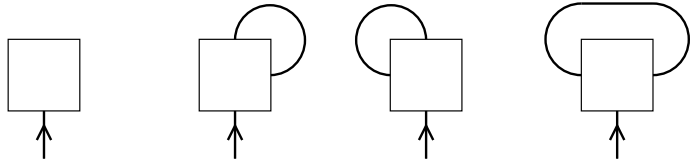}
    \caption{Loop configurations with one edge of the face $\cal
    F$
      connected to point $0$.}
    \label{fig5}
  \end{center}
\end{figure}

Now consider the contributions to the holomorphicity equation
(\ref{holo}) where $r$ is one of the edges of an elementary
square. Since the open curve ends at $r$, it must first intersect
the square on some edge $r'$ (not necessarily the same as $r$).
The configurations can then be decomposed according to the
different possible $r'$. Without loss of generality we take it to
be on the lower edge in Fig.~\ref{fig5}. The configurations can
then be further decomposed into classes corresponding to the
various cases shown in this figure. In the last three cases a
curve connects two of the other edges: it may be part of a closed
loop or of the open curve. The idea is to sum over all possible
ways of connecting up the curves internally through the square,
keeping the external configuration fixed, and to try to satisfy
(\ref{holo}) for \em each \em of these configurations. This can be
done, and yields the following linear system for the weights:
\begin{eqnarray}
  t + \mu u_1 - \mu \lambda^{-1} u_2 - v
  &=& 0 \label{eq:syst-On1} \\
  -\lambda^{-1} u_1 + n u_2 + \lambda \mu v - \mu \lambda^{-1} (w_1+n w_2)
  &=& 0 \label{eq:syst-On2} \\
  n u_1 - \lambda u_2 - \mu \lambda^{-2} v + \mu (n w_1 + w_2)
  &=& 0 \label{eq:syst-On3} \\
  - \mu \lambda^{-2} u_1 + \mu \lambda u_2 + n v - \lambda^{-2} w_1 - \lambda^2 w_2
 &=&0\,, \label{eq:syst-On4}
\end{eqnarray}
where we have set $\lambda = e^{i \pi s}\,, \varphi = (s+1)
\alpha\,, \mu = e^{i \varphi}$. For real weights,
(\ref{eq:syst-On1}--\ref{eq:syst-On4}) are four complex linear
equations for six real unknowns\\
$(t,u_1,u_2,v,w_1,w_2)$, and we have the relations:
\begin{eqnarray*}
{\rm Im}\ \left[
    (n+1) \ \{\ref{eq:syst-On1}\} - \lambda \mu^{-1} \
    \{\ref{eq:syst-On2}\}
    + \mu^{-1} \ \{\ref{eq:syst-On3}\}
  \right] &=& 0 \\
   {\rm Im}\ \left[
    \lambda \mu^{-1} (\lambda^2-n \lambda^{-2}) \
    \{\ref{eq:syst-On2}\}
    + \mu^{-1} (n \lambda^2 - \lambda^{-2}) \
    \{\ref{eq:syst-On3}\}
    - (n^2-1) \ \{\ref{eq:syst-On4}\}
  \right] &=& 0\,.
\end{eqnarray*}
Thus, we can generally reduce
(\ref{eq:syst-On1}--\ref{eq:syst-On4}) to a $6 \times 6$ real
system.

\begin{figure}
  \begin{center}
    \input{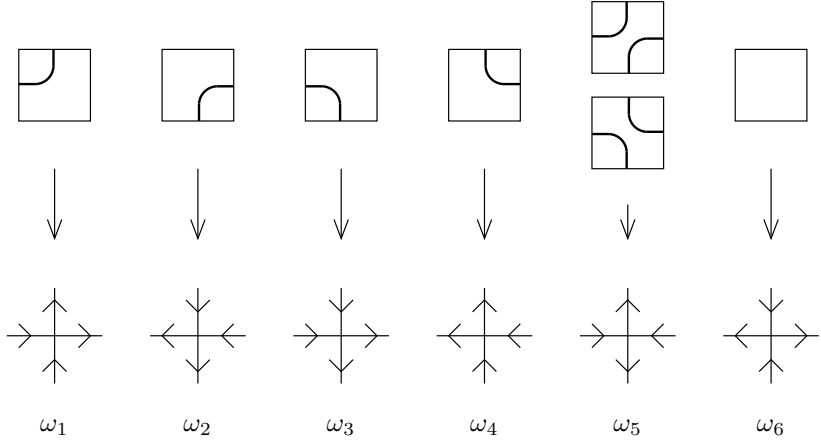}
    \caption{Mapping of the O$(n)$ model onto the six-vertex model for $n=1,v=0$.}
    \label{fig:On-6v}
  \end{center}
\end{figure}

There are two classes of solutions, for vanishing and
non-vanishing $v$. First, if $v=0$, then the configurations
corresponding to \{\ref{eq:syst-On4}\} never occur, and so this
equation does not hold. In the special case $n=1$, there exists a
non-trival solution {\it for any value of $s$}:
\begin{equation}
  t= \sin \pi s\,, \ u_1=\sin(\varphi-\pi s)\,, \ u_2=\sin \varphi\,,
  \ w_1+w_2=\sin \pi s\,.
\end{equation}
It turns out \cite{ikhlefcardy} that this model can be mapped onto
the six-vertex model (see Figure~\ref{fig:On-6v}), with weights
$\omega_1=\omega_2=\sin(\varphi-\pi s), \omega_3=\omega_4=\sin
\varphi, \omega_5=\omega_6=\sin \pi s$. The corresponding
anisotropy parameter is $\Delta = \cos \pi s$. This is an example
of a model admitting a holomorphic observable on the lattice, but
for which the scaling limit of the corresponding curve \em cannot
\em be described by simple SLE. This is because, for ordinary SLE,
the central charge of the CFT is directly related to the SLE
parameter $\kappa$ \cite{sle} and hence to the conformal spin $s$:
$c = 2s(5-8s)/(2s+1)$. In the present case, since the boundary
conditions for the six-vertex model are not twisted, its scaling
limit has central charge $c=1$ for all $\Delta$. However the
conformal spin $s$ varies continuously with $\Delta$. Therefore
the scaling limit of the curve can be SLE, with $\kappa=4$, for at
most one value (in fact $\Delta=1/\sqrt2$.) We conjecture that
other values of $\Delta$ in fact correspond to a variant of SLE
called SLE$(4,\rho)$.

For $v=0\,, n \neq -1$, we get a $5 \times 5$ linear system, with
determinant $(n^2-1)^2 \sin \varphi \ \sin (\varphi-\pi s)$.
Imposing $\sin \varphi=0$ yields $(s+1) \alpha = m \pi$ and in
turn $s=m'$, where $m,m'$ are integers. Thus, for the solution to
exist at any value of $\alpha$, we have to set $s=-1$. The
Boltzmann weights are then:
\begin{equation} \label{eq:n+1}
  t = -u_1-u_2\,, \ w_1= -u_1\,, \ w_2= -u_2\,.
\end{equation}
The solution of the case $\sin (\varphi-\pi s)=0$ is similar, and
leads to the same Boltzmann weights and spin $s=-1$. If we change
the sign of $u_1,u_2$, then the model (\ref{eq:n+1}) is equivalent
to a dense loop model which corresponds to the critical $q$-state
Potts model, with weight per loop $\sqrt{q}=n+1$. To see this,
fill empty spaces with loops of weight $1$ ({\it ghost loops}).
The local weights do not depend on the type of loops involved
(actual or ghost loops), so each loop has an overall weight $n+1$.
As a consequence, the dense loop model has a lattice
antiholomorphic observable ($s<0$), besides the holomorphic one
found in~\cite{cardyriva}. However, several arguments rule out the
hypothesis that this corresponds to an antiholomorphic field in
the continuum limit. First, $\psi_{s=-1}(z)$ is lattice
antiholomorphic for any $Q>0$, whereas it is well known that the
self-dual Potts model is only critical for $0 \leq Q \leq 4$.
Furthermore, the ratio $u_1/u_2$ in \ref{eq:n+1} does not depend
on the angle $\alpha$, which means that the same model has an
antiholomorphic observable for any deformation angle: this is not
acceptable physically in the continuum limit. So we conclude that,
in the case $v=0$ and generic $n \neq -1$, the holomorphicity
conditions (\ref{holo}) for the dilute O$(n)$ model merely lead to
the case of the dense loop model, but the corresponding
$\psi_s(z)$ is not a candidate for an antiholomorphic field in the
continuum limit.

Let us now discuss the solutions of second class ($v\neq 0$), for
a generic value of $n$. We get the $6 \times 6$ real system:
$$
  ({\rm Re} \ \{\ref{eq:syst-On1}\}, {\rm Re} \ \{\ref{eq:syst-On2}\}, {\rm Im} \ \{\ref{eq:syst-On2}\},
  {\rm Re} \ \{\ref{eq:syst-On3}\}, {\rm Im} \ \{\ref{eq:syst-On3}\}, {\rm Re} \ \{\ref{eq:syst-On4}\} )\,,
$$
with determinant: $(n^2-1) \sin \varphi \sin (\varphi-\pi s)
\left(2 \cos 4\pi s -3n+n^3 \right)$. Non-trivial solutions exist
if the spin satisfies:
\begin{equation} \label{eq:s-On1}
  \cos 4 \pi s = \cos 6 \eta\,.
\end{equation}
The various solutions to \ref{eq:s-On1} can be parameterized by
extending the range of $\eta$ to $[-\pi,\pi]$, and setting:
\begin{equation} \label{eq:s-On2}
  s = \frac{3 \eta}{2\pi} - \frac12\,.
\end{equation}
Then, we get the second class of solutions, with Boltzmann
weights:
\begin{eqnarray}
t   &=& -\sin \left( 2\varphi-{3\eta}/{2} \right)
      + \sin {5\eta}/{2} - \sin {3\eta}/{2} + \sin {\eta}/{2} \label{bw1}\\
    u_1 &=& -2 \sin \eta \cos \left( {3\eta}/{2} - \varphi \right)\label{bw2} \\
    u_2 &=& -2 \sin \eta \sin \varphi \label{bw3}\\
    v   &=& -2 \sin \varphi \cos \left( {3\eta}/{2} - \varphi \right) \label{bw4}\\
    w_1 &=& -2 \sin(\varphi-\eta) \cos \left( {3\eta}/{2} - \varphi \right) \label{bw5}\\
    w_2 &=& 2 \cos \left({\eta}/{2} - \varphi \right) \sin
    \varphi\,.\label{bw6}
\end{eqnarray}
A remarkable fact is that the weights (\ref{bw1}--\ref{bw6}) are a
solution of the Yang-Baxter equations for the O$(n)$ loop model on
the square lattice. Indeed, after a change of variables $\varphi
\to \psi+ (\pi+\eta)/4$, they coincide with the integrable weights
in~\cite{nienhuis}. So, by solving the holomorphicity
equations~(\ref{eq:syst-On1}--\ref{eq:syst-On4}) on a deformed
lattice, we recover the integrable weights.

Other more complicated loops models (for example one with
different types of loops known as the C2(1) model \cite{c21}) can
also be studied with similar results \cite{ikhlefcardy}.

\section{Conclusions and further remarks}

We have given two main examples of lattice models, the $Z_N$ model
and the O$(n)$ model on a square lattice, in which observables can
be identified whose correlators are discretely holomorphic, as
long as the weights are both critical and satisfy the Yang-Baxter
relations. This is surprising, since the holomorphicity conditions
are linear in the weights, and work for a fixed value of the
anisotropy parameter, while the Yang-Baxter relations are cubic
functional equations for the weights. While in the case of the
$Z_N$ models the connection to integrability may be understood by
generalizing the problem to an inhomogeneous Baxter lattice, this
explanation is at present missing for the loop models. While it
would be nice to elevate these observations to a more general
result connecting holomorphicity, integrability and conformal
field theory, the counter-example given in Sec.~\ref{sec2} in
which a lattice holomorphic observable apparently does not
correspond to a conformal field should warn us that there may be
subtleties. For the $Z_N$ models for larger values of $N$,
problems in a lattice identification of the value of the conformal
spin $s$ were also noted in \cite{cardyraj}.

Although it is to be hoped that the results of Smirnov
\cite{smirnov} in using these holomorphic observables to prove
that the scaling limit of lattice curves is given by SLE can be
extended to other models, the examples we have given which
correspond to CFTs with central charge $c\geq1$ show that this may
not always be the case.

Finally, it is be hoped that at some Statistical Mechanics
Conference in the future the correct extension of these ideas to
higher dimensions will be announced!

\noindent \em Acknowledgements\em. I would like to thank my
collaborators Valentina Riva, Mohammed Rajabpour and Yacine
Ikhlef, as well as Stas Smirnov and Paul Fendley, for many
discussions on this subject. This work was supported in part by
EPSRC grant EP/D050952/1.

\end{document}